\documentstyle[12pt]{article}
\input epsf
\newcounter{abc}

\newcommand{\gE}{\mbox{$\tau$}}
\newcommand{\gM}{\mbox{$h$}}
\newcommand{\oE}{\mbox{${\cal E}$}}
\newcommand{\oEc}{\mbox{${\cal E }_c$}}
\newcommand{\oM}{\mbox{${\cal M }$}}
\newcommand{\oMc}{\mbox{${\cal M }_c$}}
\newcommand{\aE}{\mbox{$a_{{\cal E}}$}}
\newcommand{\aM}{\mbox{$a_{{\cal M}}$}}
\newcommand{\lM}{\mbox{$\lambda_{{\cal M}}$}}
\newcommand{\lE}{\mbox{$\lambda_{{\cal E}}$}}

\newcommand{\ptM}{\mbox{$\tilde{p}_{{\cal M}}$}}
\newcommand{\ptE}{\mbox{$\tilde{p}_{{\cal E}}$}}
\newcommand{\ptMstar}{\mbox{$\tilde{p}_{{\cal M}}^{\star}$}}
\newcommand{\ptEstar}{\mbox{$\tilde{p}_{{\cal E}}^{\star}$}}

\newcommand{\pLM}{\mbox{$p_L({\cal M})$}}
\newcommand{\pLE}{\mbox{$p_L({\cal E})$}}

\newcommand{\mix}{\mbox{$s$}}
\newcommand{\mixp}{\mbox{$r$} }
\newcommand{\dfac}{\mbox{$\frac{1}{1-\mix \mixp}$}}
\newcommand{\pr}{\mbox{$p_L(\phi )$}}
\newcommand{\er}{\mbox{$u(\phi )$}}
\newcommand{\uc}{\mbox{$u_c$}}
\begin{document}

\title{Concentration and energy fluctuations in a \\
critical polymer mixture}

\author{M. M\"{u}ller and N. B. Wilding \\ {\small Institut f\"{u}r
Physik, Johannes Gutenberg Universit\"{a}t,}\\ {\small Staudinger Weg
7, D-55099 Mainz, Germany}}

\date{}
\setcounter{page}{0}
\maketitle

\begin{abstract}

A semi-grand-canonical Monte Carlo algorithm is employed in
conjunction with the bond fluctuation model to investigate the
critical properties of an asymmetric binary (AB) polymer mixture. By
applying the equal peak-weight criterion to the concentration
distribution, the coexistence curve separating the A-rich and B-rich
phases is identified as a function of temperature and chemical
potential. To locate the critical point of the model, the cumulant
intersection method is used. The accuracy of this approach for
determining the critical parameters of fluids is assessed.
Attention is then focused on the joint distribution function of the
critical concentration and energy, which is analysed using a
mixed-field finite-size-scaling theory that takes due account of the
lack of symmetry between the coexisting phases.  The essential Ising
character of the binary polymer critical point is confirmed by mapping
the critical scaling operator distributions onto independently known
forms appropriate to the 3D Ising universality class. In the process,
estimates are obtained for the field mixing parameters of the model
which are compared both with those yielded by a previous method, and
with the predictions of a mean field calculation.

\end{abstract}

\thispagestyle{empty}
\begin{center}
PACS numbers 64.70Ja, 05.70.Jk
\end{center}
\newpage

\section{Introduction}

The critical point of binary liquid and binary polymer mixtures, has
been a subject of abiding interest to experimentalists and theorists
alike for many years now. It is now well established that the critical
point properties of binary liquid mixtures fall into the Ising
universality class (the default for systems with short ranged
interactions and a scalar order parameter) \cite{KUMAR}. Recent
experimental studies also suggest that the same is true for polymer
mixtures \cite{BINDER1}--\cite{LIFSCHITZ}. However, since Ising-like critical
behaviour is only apparent when the correlation length far exceeds the
polymer radius of gyration $\xi \;\gg\; \langle R_g \rangle$, the
Ising regime in polymer mixtures is confined (for all but the shortest
chain lengths) to a very narrow temperature range near the critical
point. Outside this range a crossover to mean-field type behaviour is
seen.  The extent of the Ising region is predicted to narrow with
increasing molecular weight in a manner governed by the Ginsburg
criterion \cite{GINSBURG}, disappearing entirely in the limit of
infinite molecular weight. Although experimental studies of mixtures
with differing molecular weights appear to confirm qualitatively this
behaviour \cite{JANSSEN}, there are severe problems in understanding
the scaling of the so-called ``Ginsburg number'' (which marks the
centre of the crossover region and is empirically extracted from the
experimental data \cite{SCHWAHN1}) with molecular weight

Computer simulation potentially offers an additional source of
physical insight into polymer critical behaviour, complementing that
available from theory and experiment.  Unfortunately, simulations of
binary polymer mixtures are considerable more exacting in
computational terms than those of simple liquid or magnetic systems.
The difficulties stem from the problems of dealing with the extended
physical structure of polymers. In conventional canonical simulations,
this gives rise to extremely slow polymer diffusion rates, manifest in
protracted correlation times \cite{BINDER2,PAUL}. Moreover, the
canonical ensemble does not easily permit a satisfactory treatment of
concentration fluctuations, which are an essential feature of the
near-critical region in polymer mixture. In this latter regard,
semi-grand-canonical ensemble (SGCE) Monte Carlo schemes are
potentially more attractive than their canonical counterparts. In SGCE
schemes one attempts to exchange a polymer of species A for one of
species B or vice-versa, thereby permitting the concentration of the
two species to fluctuate. Owing, however, to excluded volume
restrictions, the acceptance rate for such exchanges is in general
prohibitively small, except in the restricted case of symmetric
polymer mixtures, where the molecular weights of the two coexisting
species are identical ($N_A=N_B$). All previous simulation work has
therefore focussed on these symmetric systems, mapping the phase
diagram as a function of chain length and confirming the Ising
character of the critical point
\cite{SARIBAN,DEUTSCH1,DEUTSCH2}. Tentative evidence for a crossover
from Ising to mean field behaviour away from the critical point was
also obtained \cite{DEUTSCH4}. Hitherto however, no simulation studies
of asymmetric polymer mixtures ($N_A \neq N_B$) have been reported.

Recently one of us has developed a new type of SGCE Monte Carlo method
that ameliorates somewhat the computational difficulties of dealing
with asymmetric polymer mixtures \cite{MUELLER}. The method, which is
described briefly in section~\ref{sec:SGCE}, permits the study of
mixtures of polymer species of molecular weight $N_A$ and $N_B=kN_A$,
with $k=2,3,4\cdots$. In this paper we shall employ the new method to
investigate the critical behaviour of such an asymmetric polymer
mixture. In particular we shall focus on those aspects of the critical
behaviour of asymmetric mixtures that differ from those of symmetric
mixtures. These difference are rooted in the so called `field mixing'
phenomenon, which manifests the basic lack of energetic (Ising) symmetry
between the coexisting phases of all realistic fluid systems. Although it is
expected to have no bearing on the universal properties of fluids,
field mixing does engender certain non-universal effects in
near-critical fluids. The most celebrated of these is a weak
energy-like critical singularity in the coexistence diameter
\cite{REHR,SENGERS}, the existence of which constitutes a failure for
the `law of rectilinear diameter'. As we shall demonstrate however,
field mixing has a far more legible signature in the interplay of the
near-critical energy and concentration fluctuations, which are
directly accessible to computer simulation.

In computer simulation of critical phenomena, finite-size-scaling
(FSS) techniques are of great utility in allowing one to extract
asymptotic data from simulations of finite size \cite{PRIVMAN}.  One
particularly useful tool in this context is the order parameter
distribution function \cite{BRUCE1,BINDER3,BINDER4}.  Simulation
studies of magnetic systems such as the Ising \cite{BINDER3} and
$\phi^4$ models \cite{NICOLAIDES}, demonstrate that the critical point
form of the order parameter distribution function constitutes a useful
hallmark of a university class.  Recently however, FSS techniques have
been extended to fluids by incorporating field mixing effects
\cite{BRUCE2,WILDING1}.  The resulting mixed-field FSS theory has been
successfully deployed in Monte Carlo studies of critical phenomena in
the 2D Lennard-Jones fluid \cite{WILDING1} and the 2D asymmetric
lattice gas model \cite{WILDING2}.

The present work extends this programme of field mixing studies to 3D
complex fluids with an investigation of an asymmetric polymer
mixture. The principal features of our study are as follows. We begin
by studying the order parameter (concentration) distribution as a
function of temperature and chemical potential. The measured
distribution is used in conjunction with the equal peak weight
criterion to obtain the coexistence curve of the model. Owing to the
presence of field mixing contributions to the concentration
distribution, the equal weight criterion is found to break down
near the fluid critical point. Its use to locate the coexistence curve
and critical concentration therefore results in errors, the magnitude
of which we gauge using scaling arguments.  The field mixing
component of the critical concentration distribution is then isolated
and used to obtain estimates for the field mixing parameters of the
model. These estimates are compared with the results of a mean field
calculation.

We then turn our attention to the finite-size-scaling behaviour of the
critical {\em scaling operator} distributions. This approach
generalises that of previous field mixing studies which concentrated
largely on the field mixing contribution to the order parameter
distribution function. We show that for certain choices of the
non-universal critical parameters---the critical temperature, chemical
potential and the two field mixing parameters---these operator
distributions can be mapped into close correspondence with
independently known universal forms representative of the Ising
universality class. This data collapse serves two purposes.  Firstly,
it acts as a powerful means for accurately determining the critical
point and field mixing parameters of model fluid systems. Secondly and
more generally, it serves to clarify the {\em sense} of the
universality linking the critical polymer mixture with the critical
Ising magnet.  We compare the ease and accuracy with which the
critical parameters can be determined from the data collapse of the
operator distributions, with that possible from studies of the order
parameter distribution alone.  It is argued that for critical fluids
the study of the scaling operator distributions represent the natural
extension of the order parameter distribution analysis developed for
models of the Ising symmetry.

\section{Background}
\label{sec:back}

In this section we review and extend the mixed-field
finite-size-scaling theory, placing it within the context of the
present study.

The system we consider comprises a mixture of two polymer species
which we denote $A$ and $B$, having lengths $N_A$ and $N_B$ monomers
respectively. The configurational energy $\Phi$ (which we express in
units of $k_BT$) resides in the intra-and inter molecular pairwise
interactions between monomers of the polymer chains:

\begin{equation}
\Phi(\{{\bf r}\})=\sum_{i<j=1}^{\cal N} v(|{\bf r_i}-{\bf r_j}|)
\end{equation}
where ${\cal N}=n_AN_A+n_bN_B$ with $n_A$ and $n_B$ the number of $A$ and $B$
type polymers respectively. ${\cal N}$ is therefore the total number of
monomers (of
either species), which in the present study is maintained strictly constant.
The
inter-monomer potential $v$ is assigned a square-well form
\vbox{
\begin{eqnarray}
v(r)  = & -\epsilon \hspace{1cm}& r \le r_m \\ \nonumber
v(r)  = & 0        \hspace{1cm}&  r > r_m
\end{eqnarray}
}
where $\epsilon$ is the well depth and $r_m$ denotes the maximum range
of the potential.  In accordance with previous studies of symmetric
polymer mixtures \cite{SARIBAN,DEUTSCH1}, we assign
$\epsilon\equiv\epsilon_{AA} = \epsilon_{BB} = -\epsilon_{AB} > 0$.

The independent model parameters at our disposal are the chemical
potential difference per monomer between the two species
$\Delta\mu=\mu_A-\mu_B$,
and the well depth $\epsilon$ (both in units of $k_BT$). These
quantities serve to control the observables of interest, namely the energy
density $u$ and the monomer concentrations $\phi_A$ and $\phi_B$.
Since the overall monomer density $\phi_{\cal N}=\phi_A+\phi_B$ is
fixed, however, it is sufficient to consider only one concentration
variable $\phi$, which we take as the concentration of $A$-type
monomers:

\begin{equation}
\phi \equiv \phi_A=L^{-d}n_AN_A
\label{eq:phi}
\end{equation}
The dimensionless energy density is defined as:
\begin{equation}
u=L^{-d}\epsilon^{-1}\Phi(\{{\bf r}\})
\label{eq:u}
\end{equation}
with $d=3$ in the simulations to be chronicled below.

The critical point of the model is located by critical values of the
reduced chemical potential difference $\Delta\mu_c$ and reduced
well-depth $\epsilon_c$. Deviations of $\epsilon$ and $\Delta \mu$ from their
critical values control the sizes of the two relevant scaling field
that characterise the critical behaviour. In the absence of the special
symmetry prevailing in the Ising model, one finds that the relevant
scaling fields comprise (asymptotically) {\em linear combinations} of
the well-depth and chemical potential difference \cite{REHR}:

\begin{equation}
\tau = \epsilon_c-\epsilon +s(\Delta\mu - \Delta\mu_c) \hspace{1cm} h=\Delta\mu
- \Delta\mu_c+r(\epsilon_c-\epsilon )
\label{eq:scaflds}
\end{equation}
where $\tau$ is the thermal scaling field and $h$ is the
ordering scaling field.  The parameters \mix\ and \mixp\ are
system-specific quantities controlling the degree of field mixing. In
particular $\mixp$ is identifiable as the limiting critical gradient of the
coexistence curve in
the space of $\Delta\mu$ and $\epsilon$. The role of $s$ is somewhat
less tangible; it controls the degree to which the chemical potential
features in the thermal scaling field, manifest in
the widely observed critical singularity of the coexistence curve
diameter of fluids \cite{KUMAR,SENGERS,NAKATA}.

Conjugate to the two relevant scaling fields are scaling operators
${\cal E}$ and ${ \cal M}$, which comprise linear combinations of the
concentration and energy density \cite{BRUCE2,WILDING1}:

\begin{equation}
\oM  = \dfac \left[ \phi - s u \right] \hspace{1cm} \oE  =  \dfac \left[  u  -
r \phi \right]
\label{eq:oplinks}
\end{equation}
The operator \oM\ (which is conjugate to the ordering field $h$) is
termed the ordering operator, while \oE\ (conjugate to the
thermal field) is termed the energy-like operator.  In the
special case of models of the Ising symmetry, (for which $\mix = \mixp
=0$), \oM\ is simply the magnetisation while \oE\ is the energy
density.

Near criticality, and in the limit of large system size $L$, the
probability distributions \pLM\ and \pLE\ of the operators \oM\ and
\oE\ are expected to be describable by finite-size-scaling relations
having the form \cite{BINDER3,WILDING1}:

\setcounter{abc}{1}
\begin{eqnarray}
\label{eq:opscale1}
\pLM & \simeq &\aM ^{-1}L^{d -\lambda _{{\cal M}} } \tilde{p}_{\cal M} (\aM
^{-1}L^{d -\lambda _{{\cal M}} } \delta \oM , \aM L^{\lambda_{\cal M} }\gM ,\aE
L^{\lambda _{{\cal E}} }\gE) \\
\addtocounter{equation}{-1}
\addtocounter{abc}{1}
\pLE & \simeq &\aE ^{-1}L^{d -\lambda _{{\cal E}} } \tilde{p}_{\cal E} (\aE
^{-1}L^{d -\lambda _{{\cal E}} } \delta \oE , \aM L^{\lambda_{\cal M} }\gM ,\aE
L^{\lambda _{{\cal E}} }\gE)
\label{eq:opscale2}
\end{eqnarray}
\setcounter{abc}{0}
where $\delta \oM \equiv \oM - \oMc$ and $\delta \oE \equiv \oE -
\oEc$. The functions \ptM\ and \ptE\ are predicted to be {\em
universal}, modulo the choice of boundary conditions and the
system-specific scale-factors \aM\ and \aE\ of the two relevant
fields, whose scaling indices are $\lM = d-\beta/\nu$ and $\lE=1/\nu$
respectively. Precisely at criticality ($h=\tau=0$)
equations~\ref{eq:opscale1} and ~\ref{eq:opscale2} imply
\setcounter{abc}{1}
\begin{eqnarray}
\label{eq:opscale3}
\pLM & \simeq &\aM ^{-1}L^{\beta/\nu } \tilde{p}_{\cal M}^\star (\aM
^{-1}L^{\beta/\nu} \delta \oM )\\
\addtocounter{equation}{-1}
\addtocounter{abc}{1}
\pLE & \simeq &\aE ^{-1}L^{(1-\alpha)/\nu } \tilde{p}_{\cal E}^\star (\aE
^{-1}L^{(1-\alpha)/\nu } \delta \oE )
\label{eq:opscale4}
\end{eqnarray}
\setcounter{abc}{0}
where

\begin{equation}
\tilde{p}_{\cal M}^\star (x) \equiv  \tilde{p}_{\cal M} (x,y=0,z=0)
\hspace{1.5cm}\tilde{p}_{\cal E}^\star (x) \equiv  \tilde{p}_{\cal E}
(x,y=0,z=0)
\label{eq:critops}
\end{equation}
are functions describing the universal and statistically
scale-invariant fluctuation spectra of the scaling operators,
characteristic of the critical point.

The claim that the binary polymer critical point belongs to the Ising
universality class is expressed in its fullest form by the requirement
that the critical distribution of the fluid scaling operators \pLM\
and \pLE\ match quantitatively their respective counterparts---the
magnetisation and energy distributions---in the canonical ensemble of
the critical Ising magnet.  As we shall demonstrate, these mappings
also permit a straightforward and accurate determination of the values
of the field mixing parameters \mix\ and \mixp\ of the model.

An alternative route to obtaining estimates of the field mixing
parameters is via the field mixing correction to the order parameter
(i.e. concentration) distribution $p_L(\phi)$. At criticality, this
distribution takes the form \cite{WILDING1,WILDING2}:

\begin{equation}
\pr \simeq \aM ^{-1}L^{\beta/\nu } \left[\ptMstar(x) - s\aE
\aM ^{-1} L^{-(1-\alpha-\beta)/\nu}\frac{\partial}{\partial x}
\left( \ptMstar (x) \tilde{\omega}^{\star}(x)\right) +
O(s^2) \right]_{x=\aM ^{-1}L^{\beta/\nu } [\phi - \phi _c ] }
\label{eq:prediction}
\end{equation}
where
\begin{equation}
\tilde{\omega}^{\star }(x) =  \aE^{-1}L^{d-1/\nu}[\langle\er \rangle - \uc  -
\mixp (\phi -\phi_c )] + O(s)
\label{eq:enfunc}
\end{equation}
which we term the energy function, is a universal function
characterising the critical energy-like
operator. Equation~\ref{eq:prediction} states that to leading order in
the field mixing parameter \mix , the order parameter distribution is
a sum of two distinct universal components.  The first of these, $\ptMstar
(x)$, is simply the universal ordering operator distribution featuring
in equation~\ref{eq:opscale3}.  The second, $\frac{\partial}{\partial
x} \left( \ptMstar (x) \tilde{\omega}^{\star}(x)\right)$, is a
function characterising the mixing of the critical energy-like
operator into the order parameter distribution.  This field mixing
term is down on the first term by a factor $L^{-(1-\alpha
-\beta)/\nu}$ and therefore represents a {\em correction} to the large
$L$ limiting behaviour.  Given further the symmetries of
$\tilde{\omega }(x)$ and $\ptMstar (x)$, both of which are even
(symmetric) in the scaling variable $x$ \cite{WILDING1}, the field
mixing correction is the leading antisymmetric contribution to the
concentration distribution.  Accordingly, it can be isolated from
measurements of the critical concentration distribution simply by
antisymmetrising around $\phi_c=\langle\phi \rangle_c$. The values of \mix\ and
\mixp\ are then obtainable by matching the measured critical function
$-\mix\frac{\partial }{\partial \phi}\left \{ \pr \left [\langle\er \rangle -
\uc
- \mixp (\phi -\phi _c ) \right ] \right \}$ to the measured
antisymmetric component of the critical concentration distribution
\cite{WILDING2}. In the simulations described below we shall compare
the values of \mix\ and \mixp\ obtained by this method with those
obtained by matching the fluid operator distributions to their Ising
equivalent forms.

Finally in this section, we turn to a consideration of the
finite-size-scaling behaviour of the energy distribution function. In
contrast to the situation for the order parameter distribution
described above, it transpires that field mixing radically alters the
{\em limiting} form of the critical energy density distribution.  To
substantiate this claim we reexpress $u$ in terms of the scaling
operators. Appealing to equation~\ref{eq:oplinks}, one finds

\begin{equation}
u=\oE+\mixp\oM
\label{eq:umix}
\end{equation}
so that the critical energy density distribution is

\begin{equation}
p_L(u)=p_L (\oE +r\oM )
\label{eq:pu}
\end{equation}
Now the structure of the scaling forms~\ref{eq:opscale3}
and~\ref{eq:opscale4} show that the typical size of the fluctuations
in the energy-like operator will vary with system size like $\delta\oE\sim
L^{-(1-\alpha)/\nu}$, while the typical size of the fluctuations in the
ordering operator vary like $\delta\oM\sim L^{-\beta/\nu}$ .
Given that in general $\alpha <\beta$, it follows that
asymptotically, the contribution of $\oE$ to the argument on the
right hand side of equation~\ref{eq:pu} can be neglected, so that

\begin{equation}
p_L(u)\simeq p_L (r\oM )\simeq\aM ^{-1}\mixp L^{\beta/\nu }\tilde{p}_{\cal
M}^\star (\aM ^{-1}\mixp L^{\beta/\nu} \delta \oM )\\
\label{eq:limu}
\end{equation}
We conclude then that for sufficiently large $L$, the distribution of
the fluid critical energy density has the same functional form as the
distribution of the critical ordering operator \ptMstar\ . Given
further that \ptMstar\ possesses a {\em symmetric double-peaked} form,
while the critical energy distribution in the Ising model
$p_L(u)=\ptEstar$ possesses an {\em asymmetric single-peaked} form,
the profound influence of field mixing on the critical energy
distribution of fluids should be apparent.

\section{Monte Carlo studies}

\subsection{Algorithmic and computational aspects}
\setcounter{equation}{0}
\label{sec:SGCE}

The polymer model studied in this paper is the bond-fluctuation model
(BFM). The BFM is a coarse-grained lattice-based model that combines
computational tractability with the important qualitative features of
real polymers: monomer excluded volume, monomer connectivity and short
range interactions. Within the framework of the model, each monomer
occupies a whole unit cell of a 3D periodic simple cubic lattice.
Neighbouring monomers along the polymer chains are connected via one
of $108$ possible bond vectors. These bond vectors provide for a total
of $5$ different bond lengths and $87$ different bond angles. Thermal
interactions are catered for by a short range inter-monomer
potential. Further details concerning the model can be found in
reference \cite{BFM}.

The system we have studied comprises two polymer species $A$ and $B$
having chain lengths $N_A$ and $N_B$, with $N_B=kN_A$. The SGCE scheme
whereby polymers of type A are transformed into polymers of type B (or
vice versa) is described in appendix 1 and in reference~\cite{MUELLER}, but in
general terms operates as follows. Using a Metropolis algorithm, an
A-type polymer is formed simply by cutting a B-type polymer into $k$
equal segments.  Conversely, a B-type polymer is manufactured by
connecting together the ends of $k$ A-type polymers.  This latter
operation is, of course, subject to condition that the connected ends
satisfy the bond restrictions of the BFM. Consequently it represents
the limiting factor for the efficiency of the method, since for large
values of $k$ and $N_A$, the probability that $k$ polymer ends
simultaneously satisfy the bond restrictions becomes prohibitively
small. The acceptance rate for SGCE moves is also further reduced by
factors necessary to ensure that detailed balance is satisfied. In
view of this we have chosen $k=3, N_A=10$ for the simulations
described below, resulting in an acceptance rate for SGCE moves of
approximately $14\%$.

In addition to the compositional fluctuations associated with SGCE
moves, it is also necessary to relax the polymer configurations at
constant composition. This is facilitated by monomer moves which can
be either of the local displacement form, or of the reptation
(`slithering snake') variety \cite{BINDER1}. These moves were employed
in conjunction with SGCE moves, in the following ratios :

\begin{center}
local displacement : reptation : semi-grandcanonical = 4 : 12 : 1
\end{center}
the choice of which was found empirically to relax the configurational and
compositional modes of the system on approximately equal time scales.

In the course of the simulations, a total of five system sizes were
studied having linear extent $L=32,40,50,64$ and $80$.  An overall
monomer filling fraction of $8\phi_{\cal N}=0.5$ was chosen,
representative of a dense polymer melt \cite{PAUL}. Here the factor of
$8$ constitutes the monomeric volume, each monomer occupying $8$ lattice
sites. The cutoff range of the inter-monomeric square well potential
was set at $r_m=\sqrt{6}$ (in units of the lattice spacing), a choice
which ensures that the first peak of the correlation function is
encompassed within the range of the potential. The observables sampled
in the simulations were the A-monomer concentration $\phi$ and the
energy density $u$ (cf.  equations~\ref{eq:phi} and~\ref{eq:u}). The
joint distribution $p_L(\phi,u)$ of these quantities was accumulated
in the form of a histogram, with successive samples being separated by
$12.5$ semi-grand canonical sweeps in order to reduce
correlations. The final histograms for the lattice sizes $L=40$ and
$64$ each comprised some $5 \times 10^5$ entries. Assistance in
exploring the phase space of the model, was provided by means of the
multi-histogram reweighting technique \cite{FERRENBERG,DEUTSCH3}. This
technique allows one to generate estimated histograms $p_L(\phi,u)$
for values of the control parameters $\epsilon$ and $\Delta\mu$ other
than those at which the simulations were actually performed.  Such
extrapolations are generally very reliable in the neighbourhood of the
critical point, due to the large critical fluctuations
\cite{FERRENBERG}.

\subsection{The coexistence curve and critical limit}
\label{sec:cl1}

In general for fluid systems, the coexistence curve is not known {\em
a-priori} and must therefore be identified empirically as a prelude to
locating the critical point itself. One computational criterion that
can be used to effect this identification is the so-called `equal
weight criterion' for the order parameter (concentration) distribution
function $p_L(\phi)=\int du p_L(\phi,u)$ \cite{EWR}.  Precisely on
coexistence and for temperatures well below criticality, $p_L(\phi)$
will comprise two well-separated gaussian peaks of equal weight, but
unequal heights and widths. The centres of these peaks identify the
concentrations of the coexisting A-rich and A-poor phases. For a given
subcritical well-depth $\epsilon$, the coexistence value for the
chemical potential difference $\Delta\mu_{cx}$ can therefore be
obtained by adjusting $\Delta\mu$ until the concentration distribution
satisfies the condition:

\begin{equation}
\int_0^{\phi^*} p_L(\phi,\epsilon,\Delta \mu_{cx})
\;d\phi = \int_{\phi^*}^{\phi_{\cal N}} p_L(\phi,\epsilon, \Delta
\mu_{cx}) \;d\phi \label{eq:ewr}
\end{equation}
where $\phi^*$ is a parameter defining the boundary between the two peaks.

Well below criticality, the value of $\Delta\mu_{cx}$ obtained from
the equal weight criterion is insensitive to the choice of $\phi^*$,
provided it is taken to lie approximately midway between the peaks and
well away from the tails.  As criticality is approached however, the
tails of the two peaks progressively overlap making it impossible to
unambiguously define a peak in the manner expressed by
equation~\ref{eq:ewr}. For models of the Ising symmetry, for which the
peaks are symmetric about the coexistence concentration $\phi_{cx}$,
the correct value of $\Delta\mu_{cx}$ can nevertheless be obtained by
choosing $\phi^*= \langle\phi\rangle$ in equation~\ref{eq:ewr}. In
near-critical fluids, however, the imposed equal weight rule forces a
shift in the chemical potential away from its coexistence value in
order to compensate for the presence of the field mixing
component. Only in the limit as $L\rightarrow \infty$ (where the field
mixing component dies away), will the critical order parameter
distribution be symmetric allowing one to choose
$\phi^*=\langle\phi\rangle$ and still obtain the correct coexistence
chemical potential. Thus for finite-size systems, use of the equal
weight criterion is expected to lead to errors in the determination of
$\Delta\mu_{cx}$ near the critical point.  Although this error is much
smaller than the uncertainty in the location of the critical point
{\em along} the coexistence curve (see below), it can lead to
significant errors in estimates of the critical concentration
$\phi_c$.

To quantify the error in $\phi_c$ it is necessary to match the
magnitude of the field mixing component of the concentration
distribution $w(\delta p_L)$, to the magnitude of the peak weight
asymmetry $w^\prime(\delta\mu)$ associated with small departures
$\delta\mu = \Delta\mu-\Delta\mu_{cx}$ from coexistence:

\begin{equation}
w(\delta p_L) =  w^\prime(\delta\mu)
\end{equation}
Now from equation~\ref{eq:prediction}

\begin{equation}
w(\delta p_L) \approx \int_{\phi^*}^{\phi_{\cal N}} d\phi\; \delta
p_L(\phi)\sim L^{-(1-\alpha-\beta)/\nu}
\end{equation}
while
\begin{equation}
w^\prime(\delta\mu) \approx \int_{\phi^*}^{\phi_{\cal N}} \; d \phi \;
\frac{\partial p_L(\phi)}{\partial \Delta \mu} \delta \mu
\propto L^{(\beta+\gamma)/\nu} \;\; \delta \mu
\end{equation}
It follows that the error in $\Delta \mu_{cx}$ varies with system size like:

\begin{equation}
\delta \mu \propto L^{-(1-\alpha+\gamma)/\nu}
\label{eq:shiftmu}
\end{equation}
Accordingly the error in the critical concentration
(obtained as the first moment of the concentration distribution)
varies with system size like
\begin{equation}
\delta\phi_c=\chi (L)\delta\mu=L^{\gamma/\nu}\delta\mu\sim
L^{-(1-\alpha)/\nu}
\label{eq:shiftphi}
\end{equation}
Note also that an analogous treatment of the shift in the auxiliary variable
$\phi^*$ leads to the same $L$-dependence.

Measurements of the concentration distribution were performed in
conjunction with the equal weight criterion, to locate the coexistence
curve as a function of well depth and chemical potential. The results
are shown is figure~\ref{fig:cxcurve}. Since in finite-size systems,
the order parameter distribution exhibits a double peaked structure
even above the critical temperature, the data shown also represent the
analytic continuation of the true coexistence curve that persists in
finite-size systems \cite{WILDING1}.  To determine the position of the
critical point on this line of pseudo coexistence, the cumulant
intersection method was employed.  The fourth order cumulant ratio
$G_L$ is a dimensionless quantity that characterises the form of a
function. It is defined as

\begin{equation}
G_L=1-\frac{\langle m^4 \rangle}{3\langle m^2 \rangle^2}
\end{equation}
where $m^2$ and $m^4$ are the second and fourth moments respectively
of the order parameter $m=\phi-\langle\phi\rangle$. To the extent that
field mixing corrections can be neglected, the critical order
parameter distribution function is expected to assume a universal
scale invariant form. Accordingly, when plotted as a function of
$\epsilon$, the coexistence values of $G_L$ for different system sizes
are expected to intersect at the critical well depth
$\epsilon_c$ \cite{BINDER4}. This method is particularly attractive for
locating the
critical point in fluid systems because the even moments of the order
parameter distribution are insensitive to the antisymmetric (odd)
field mixing contribution.  Figure~\ref{fig:cumulant} displays the
results of performing this cumulant analysis. A
well-defined intersection point occurs for a value $G_L=0.47$, in
accord with previously published values for the 3D Ising universality
class \cite{CUMULANT}. The corresponding estimates for the critical
well depth and critical chemical potential are

\[
\epsilon_c=0.02756(15) \hspace{1cm} \Delta\mu_c = 0.003603(15)
\]

It is important in this context, that a distinction be drawn between
the errors on the location of the critical point, and the error with
which the coexistence curve can be determined. The uncertainty in the
position of the critical point {\em along} the coexistence curve, as
determined from the cumulant intersection method, is in general
considerably greater than the uncertainty in the location of the
coexistence curve itself. This is because the order parameter
distribution function is much more sensitive to small
deviations off coexistence (due to finite $\epsilon-\epsilon_{cx}$ or
finite $\Delta\mu-\Delta\mu_{cx}$) than it is for deviations along the
coexistence curve, ($\epsilon$ and $\Delta\mu$ tuned together to
maintain equal weights). In the present case, we find that the errors
on $\Delta\mu_c$ and $\epsilon_c$ are approximately $10$ times those of
the coexistence values $\epsilon_{cx}$ and $\Delta\mu_{cx}$ near the
critical point.

The concentration distribution function at the assigned value of
$\epsilon_c$ and the corresponding value of $\Delta\mu_{cx}$,
(determined according to the equal weight rule with $\phi^*=<\phi>$),
is shown in figure \ref{fig:ewr} for the $L=40$ and $L=64$ system
sizes. Also shown in the figure is the critical magnetisation
distribution function of the $3D$ Ising model obtained in a separate
study \cite{HILFER}. Clearly the $L=40$ and $L=64$ data differ from
one another and from the limiting Ising form. These discrepancies
manifest both the pure antisymmetric field mixing component of the
true (finite-size) critical concentration distribution, and small
departures from coexistence associated with the inability of the equal
weight rule to correctly identify the coexistence chemical
potential. To extract the infinite-volume value of $\phi_c$ from the
finite-size data, it is therefore necessary to extrapolate to the
thermodynamic limit.  To this end, and in accordance with
equation~\ref{eq:shiftphi}, we have plotted $\phi_{cx}(L)$,
representing the first moment of the concentration distribution
determined according to equal weight criterion at the assigned value
of $\epsilon_c$, against $L^{(1-\alpha)/\nu}$ . This extrapolation
(figure~\ref{fig:phic}) yields the infinite-volume estimate:

\[
\phi_c = 0.03813(19)
\]
corresponding to a reduced A-monomer density $\phi_c/\phi_{\cal N} =
0.610(3)$. The finite-size shift in the value of $\phi_{cx}(L)$ is of
order $2\%$.

We turn next to the determination of the field mixing parameters $r$
and $s$. The value of $r$ represents the limiting critical gradient of
the coexistence curve which, to a good approximation, can be simply
read off from figure~\ref{fig:cxcurve} with the result $r =
-0.97(3)$. Alternatively (and as detailed in \cite{WILDING1}) $r$ may
be obtained as the gradient of the line tangent to the measured critical
energy function (equation~\ref{eq:enfunc}) at $\phi=\phi_c$. Carrying
out this procedure yields $r = -1.04(6)$.

The procedure for extracting the value of the field mixing parameter
\mix\ from the concentration distribution is rather more involved, and
has been described in detail elsewhere \cite{WILDING1,WILDING2}.  The
basic strategy is to choose \mix\ such as to satisfy \[ \delta
p_L(\phi)=-\mix\frac{\partial }{\partial \phi}\left \{ \pr \left
[\langle\er \rangle - \uc - \mixp (\phi -\phi _c) \right ] \right \}
\] where $\delta p_L(\phi)$ is the measured antisymmetric field mixing
component of the critical concentration distribution \cite{WILDING2},
obtained by antisymmetrising the concentration distribution about
$\phi_c(L)$ and subtracting additional corrections associated with
small departures from coexistence resulting from the failure of the
equal weight rule. Carrying out this procedure for the $L=40$ and
$L=64$ critical concentration distributions yields the field mixing
components shown in figure~\ref{fig:asym}. The associated estimate for
\mix\ is $0.06(1)$. Also shown in figure~\ref{fig:asym} (solid line)
is the predicted universal form of the 3D order parameter field mixing
correction $-\frac{\partial}{\partial x} \left( \ptMstar (x)
\tilde{\omega}^{\star}(x)\right)$ (cf. equation~\ref{eq:prediction})
obtained from independent Ising model studies \cite{HILFER}.  Clearly
the measured functional form of the field mixing correction is in
reasonable accord with the universal prediction. We attribute the
residual discrepancies to field mixing contributions of order $s^2$ or
higher, not included in equation~\ref{eq:prediction}.  The directions
of the two relevant scaling fields \gM\ and \gE\ corresponding to the
measured values of \mix\ and \mixp\ are indicated on
figure~\ref{fig:cxcurve}.

\subsection{The critical limit revisited : Scaling operator distributions}

In this subsection we consider an alternative method for locating the
critical point and determining the field mixing parameters \mix\ and
\mixp, that circumvents some of the difficulties associated with using
the order parameter distribution alone.

The method focuses on the ordering and energy-like scaling operator
distributions \pLM\ and \pLE\ (cf.  equations~ \ref{eq:opscale3},
\ref{eq:opscale4}), which are obtained from the joint distribution of
the energy density and concentration $p_L(\phi,u)$ as

\begin{equation}
\pLM=p_L\left (\frac{\phi-\mix u}{1-sr}\right ) \hspace{1cm}
\pLE=p_L\left (\frac{u-\mixp \phi}{1-sr} \right )
\end{equation}

Precisely at criticality, \pLM\ is expected to match the universal
fixed point function \ptMstar. This suggests that the critical
parameters can be readily located by simultaneously adjusting \mix
,$\epsilon$ and $\Delta\mu$ until $p_L([\phi-\mix u]/(1-sr))$ matches
\ptMstar\ (modulo the choice of non-universal scale parameters
implicit in the definition of the scaling variable).  The results of
performing this procedure for \pLM\ are displayed in
figure~\ref{fig:op} for the $L=40$ and $L=64$ system sizes. The
quality of the data collapse lends substantial support to the
contention that the binary polymer critical point does indeed belong
to the Ising universality class. The corresponding values of the
critical parameters $\epsilon, \Delta\mu_c$ and $\mix$ are

\begin{equation}
\epsilon_c=0.02756(15) \hspace{1cm} \Delta\mu_c = 0.003603(15)
\hspace{1cm}\mix=0.06(1)
\label{eq:critpt}
\end{equation}
in good agreement with those determined previously. We note however
that the present method permits the determination of \mix\ without the
need to isolate the field mixing component of the concentration
distribution, a procedure that is somewhat cumbersome and which is anyway
only accurate to leading order in \mix\ \cite{WILDING2}.

The value of the field mixing parameter \mixp, is intimately
associated with the critical energy distribution $p_L(u)$, the form of
which is shown in figure~\ref{fig:en} for both the $L=40$ and $L=64$
system sizes. The corresponding mapping of the scaling operator
distribution function $p_L([u-\mixp\phi]/(1-sr))$ onto the universal
energy distribution of the 3D Ising model, \ptEstar\ is shown in
figure~\ref{fig:enop}.  Again the agreement with the universal form is
gratifying, although there are small discrepancies which we attribute
to corrections to scaling. The data collapse implies a value
$\mixp=-1.00(3)$, which also agrees to within error with the value
obtained previously.

With regard to the critical energy distributions of
figure~\ref{fig:en}, we note that the distributions are not at all
reminiscent of the critical Ising energy distribution of
figure~\ref{fig:enop}. Neither, however, are they similar to
\ptMstar\ (cf. figure~\ref{fig:op}), which it was claimed they
match in the limit $L\rightarrow \infty$
(c.f. section~\ref{sec:back}). This discrepancy implies that the
system size is still too small to reveal the asymptotic behaviour
Nevertheless the data do afford a test of the {\em approach} to
the limiting regime, via the FSS behaviour of the variance of the
energy distribution. Recalling equation~\ref{eq:limu}, we anticipate
that this variance exhibits the same FSS behaviour as
the Ising {\em susceptibility}, namely:

\begin{equation}
L^d(\langle u^2\rangle-u_c^2)\sim L^{\gamma/\nu}
\end{equation}
By contrast, the variance of the scaling operator \oE\ is expected to
display the FSS behaviour of the Ising specific heat:

\begin{equation}
L^d(\langle\oE^2\rangle-\oE_c^2)\sim L^{\alpha/\nu}.
\end{equation}
Figure~\ref{fig:cv} shows the measured system size dependence of these
two quantities at criticality. Also shown is the scaled variance of
the ordering operator $L^d(\langle\oM^2\rangle-\oM_c^2)\sim
L^{\gamma/\nu}.$ Straight lines of the form $L^{\gamma/\nu}$ and
$L^{\alpha/\nu}$, (indicative of the FSS behaviour of the Ising
susceptibility and specific heat respectively) have also been
superimposed on the data. Clearly for large $L$, the scaling behaviour
of the variance of the energy distribution does indeed appear to
approach that of the ordering operator distribution.

\section{Mean field calculations}
\setcounter{equation}{0}

In this section we derive approximate formulae for the values of the
field mixing parameters \mix\ and \mixp\ on the basis of a mean field
calculation.

Within the well-known Flory-Huggins theory of polymer mixtures, the
mean-field equation of state takes the form:

\begin{equation}
\Delta\mu = \frac{1}{N_A} \ln (\rho) + \frac{1}{N_B} \ln (1-\rho) - 2z\epsilon
(2 \rho -1) +C
\label{eq:eqofst}
\end{equation}
In this equation, $z\approx 2.7$ is the effective monomer coordination
number, whose value we have obtained from the measured pair
correlation function.  $\rho=\phi/\phi_{\cal N}$ is the
density of A-type monomers and the constant $C$ is the entropy density
difference of the pure phases, which is independent of temperature and
composition. In what follows we reexpress $\rho$ by the concentration $\phi$.

The critical point is defined by the condition:

\begin{equation}
\frac{\partial}{\partial \phi} \Delta\mu_c =
\frac{\partial ^2}{\partial \phi ^2} \Delta\mu_c = 0
\end{equation}
where $\Delta \mu_c = \Delta \mu(\phi_c,\epsilon_c)$.
This relation can be used to determine the critical concentration
and critical well-depth, for which one finds

\begin{equation}
\frac{\phi_c}{\phi_{\cal N}} = \frac{1}{1+1/\sqrt{k}} \qquad \mbox{and } \qquad
\frac{1}{\epsilon_c} = z \frac{4 N_A N_B}{(\sqrt{N_A}+\sqrt{N_B})^2}
\label{eq:critpars}
\end{equation}

Below the critical point a first order phase transition occurs between
the A-rich and A-poor phases. To determine the location of the phase
boundary we employ a Landau expansion of the equation of state in
terms of the small parameters $\delta \phi = \phi -\phi_c$ and $\delta
\epsilon = \epsilon - \epsilon_c$:

\begin{equation}
\Delta \mu = \Delta \mu_c + r' \delta \epsilon - a \delta
\epsilon \delta \phi + b \delta \phi^3 + c \delta \phi^4 + \cdots
\end{equation}
where the expansion coefficients take the form
\begin{equation}
r^\prime= -2z(2 \frac{\phi_c}{\phi_{\cal N}} -1 ) \hspace{1cm} a =
\frac{4z}{\phi_{\cal N}}
\hspace{1cm} b = \frac{(1+\sqrt{k})^4}{3\sqrt{k}N_A \phi_{\cal N}^3}
\hspace{1cm} c=\frac{(k-1)(1+\sqrt{k})^4}{4 k N_A \phi_{\cal N}^4}
\end{equation}
The phase boundary itself is specified by the binodal condition

\begin{equation}
\Delta \mu_{cx}(\epsilon) =  \Delta \mu (\phi_+,\epsilon) = \Delta \mu
(\phi_-,\epsilon) =
\frac{\int_{\phi_-}^{\phi_+}d\phi \; \Delta \mu(\phi,\epsilon)  }{\phi_+ -
\phi_-}
\end{equation}
where $\phi_-$ and $\phi_+$ denote the concentration of A monomers in the
A-poor phase and
A-rich phases respectively. Thus to leading order in $\epsilon$, the phase
boundary is given by :

\begin{equation}
\Delta \mu_{cx}(\epsilon) = \Delta \mu_c + r' \delta \epsilon_c + \cdots
\label{eq:coex}
\end{equation}
Consequently we can identify the expansion coefficient $r^\prime$ with
the field mixing parameter $r$ (c.f. equation~\ref{eq:scaflds}) that controls
the
limiting critical gradient of the coexistence curve in the space of
$\Delta\mu$ and $\epsilon$. Substituting for $\Delta\mu_c$ and
$\epsilon_c$  in equation~\ref{eq:coex} and setting $k=3$, we find
$r=-1.45$, in order-of-magnitude agreement with the FSS analysis of
the simulation data.

In order to calculate the value of the field mixing parameter \mix ,
it is necessary to obtain the concentration and energy densities of
the coexisting phases near the critical point. The concentration of
A-type monomers in each phase is given by

\begin{equation}
\delta \phi_{\pm} = \phi_{\pm} -\phi_c = \pm \sqrt{\frac{a \delta
\epsilon}{b}} - \frac{2ac \delta \epsilon}{5 b^2} + \cdots
\end{equation}
so that the variation of the order parameter along the coexistence curve is:

\begin{equation}
\langle \delta \phi \rangle = \left\{ \frac{\phi_+ +
\phi_-}{2} - \phi_c \right\} = -\frac{2 ac \delta
\epsilon}{5 b^2} \label{eq:phi_coex}
\end {equation}
A similar calculation for the energy density yields:

\begin{equation}
\langle u(\phi)\rangle = -\frac{\phi_{\cal N}}{2} \left(
z_s+z(2\rho-1)^2\right) = -\frac{\phi_{\cal N}}{2} \left(
z_s+z(2\frac{\phi_c}{\phi_{\cal N}}-1)^2\right) + r\delta \phi -
\frac{2z}{\phi_{\cal N}} \delta \phi^2
\end{equation}
where $z_s$ is the coordination number of the intra-chain thermal
interactions. The variation of the energy density along the
coexistence curve then follows as:

\begin{equation}
\langle \delta u \rangle = \frac{u(\phi_+)+u(\phi_-)}{2} - u(\phi_c) =
-\frac{2za}{ b \phi_{\cal N}} \left( 1 + r \frac{c \phi_{\cal N}}{5zb}
\right)\delta \epsilon + \cdots
\label{eq:u_coex}
\end{equation}
Now since $(1-rs)\langle \oM \rangle = \langle \delta \phi \rangle -s
\langle \delta u \rangle$ vanishes along the coexistence line,
equations \ref{eq:phi_coex} and \ref{eq:u_coex} yield the following
estimate for the field mixing parameter $s$:

\begin{equation}
s = \frac{\langle \delta \phi \rangle}{\langle \delta u \rangle} =
\frac{c \phi_{\cal N}}{5zb \left( 1 + r \frac{c \phi_{\cal N}}{5zb}\right)} =
\frac{3(k-1)}{20 z
\sqrt{k} \left(1 + r \frac{3(k-1)}{20 z \sqrt{k}} \right)}
\label{eq:mfs}
\end{equation}

Thus within the mean field framework, the field mixing parameter $s$
is controlled by the ratio of the fifth and fourth order coefficient
of the Landau expansion of the free energy. For the present case ($
k=3$) equation~\ref{eq:mfs} yields $\mix=0.070$, in good agreement
with the value obtained from the FSS analysis of the simulation
data. It is also similar in magnitude to the values of $\mix$ measured
for the 2D Lennard-Jones fluid \cite{WILDING1} and 2D asymmetric
lattice gas model \cite{WILDING2}.  The sign of the product $rs$
differs however from that found at the liquid-vapour critical
point. In the present context this product is given by

\begin{equation}
\frac{rs}{1-rs} = -\frac{3}{10} \frac{(\sqrt{N_A}-\sqrt{N_B})^2}{\sqrt{N_A
N_B}}
\end{equation}
However an analogous treatment of the van der Waals fluid predicts a
positive sign $rs$, in agreement with that found at the liquid vapour
critical point \cite{WILDING1,WILDING2}.

\section{Concluding remarks}

In summary we have employed a semi-grand-canonical Monte Carlo
algorithm to explore the critical point behaviour of a binary polymer
mixture.  The near-critical concentration and scaling operator
distributions have been analysed within the framework of a mixed-field
finite-size scaling theory. The scaling operator distributions were
found to match independently known universal forms, thereby confirming
the essential Ising character of the binary polymer critical point.
Interestingly, this universal behaviour sets in on remarkably short
length scales, being already evident in systems of linear extent
$L=32$, containing only an average of approximately $100$ polymers.

Regarding the specific computational issues raised by our study, we
find that the concentration distribution can be employed in
conjunction with the cumulant intersection method and the equal weight
rule to obtain a rather accurate estimate for the critical temperature
and chemical potential. The accuracy of this estimate is not adversely
affected by the antisymmetric (odd) field mixing contribution to the
order parameter distribution, since only even moments of the
distribution feature in the cumulant ratio. Unfortunately, the method
can lead to significant errors in estimates of the critical
concentration $\phi_c$, which {\em are} sensitive to the magnitude of
the field mixing contribution. The infinite-volume value of $\phi_c$
must therefore be estimated by extrapolating the finite-size data to
the thermodynamic limit (where the field mixing component
vanishes). Estimates of the field mixing parameters \mix\ and \mixp\
can also be extracted from the field mixing component of the order
parameter distribution, although in practice we find that they can be
determined more accurately and straightforwardly from the data
collapse of the scaling operators onto their universal fixed point
forms.

In addition to clarifying the universal aspects of the binary polymer
critical point, the results of this study also serve more generally to
underline the crucial role of field mixing in the behaviour of
critical fluids. This is exhibited most strikingly in the form of the
critical energy distribution, which in contrast to models of the Ising
symmetry, is doubly peaked with variance controlled by the Ising
susceptibility exponent. Clearly therefore close attention must be
paid to field mixing effects if one wishes to perform a comprehensive
simulation study of critical fluids. In this regard, the scaling
operator distributions are likely to prove themselves of considerable
utility in future simulation studies. These operator distributions
represent the natural extension to fluids of the order parameter
distribution analysis deployed so successfully in critical phenomena
studies of (Ising) magnetic systems. Provided therefore that one works
within an ensemble that affords adequate sampling of the near-critical
fluctuations, use of the operator distribution functions should also
permit detailed studies of fluid critical behaviour.

\subsection*{Acknowledgements}

The authors thank K. Binder for helpful discussions.  NBW acknowledges
the financial support of a Max Planck fellowship from the Max
Planck Institut f\"{u}r Polymerforschung, Mainz. Part of the
simulations described here were performed on the KSR1 parallel
computer at the Universit\"{a}t Mannheim and on the CRAY Y-MP at the
HLRZ J\"{u}lich. Partial support from the Deutsche
Forschungsgemeinschaft (DFG) under grant number Bi314/3-2 is also
gratefully acknowledged.

\subsection*{Appendix A}

In this appendix we give a brief description of our
semi-grandcanonical algorithm for polymer mixtures with chain lengths
$N_B=k N_A$,$k=3$. A more detailed presentation of the method can be
found in reference \cite{MUELLER}.

As illustrated in figure \ref{fig:algo} the SGCE Monte-Carlo moves
consists in either joining together $k$ A-polymers to form a B chain,
or alternatively, cutting a B-polymer into $k$ equal segments, each of
which is then an $A$ chain.  The $B$ chain formation step is attempted
with probability $n_A/(n_A+n_B)$, and proceeds as follows. First one
starts by choosing a random A-polymer end (a given end is selected
with probability $1/2n_A$). One then determines the number $\nu_1$ of
neighbouring A-ends that satisfy the bonding constraints. Of the
$\nu_1$ possible ends to which a bond might be formed, one is chosen
randomly and the ends connected together. In the same way one computes
the number $\nu_2$ of possible bonding partners for the remaining end
of the second A-polymer and makes a connection if possible.  Thus the
proposition probability for B polymer formation is given by:
$P^{prop}_{kA\to B'} = 1/2(n_A+n_B)\nu_1\nu_2$.  Finally the move is
accepted with probability \[P^{acc}_{kA \to B'} =
\min(1,\exp(-\beta\Delta E(kA\to B')-N_B\Delta\mu)\]

The formation procedure for A chains simply involves cutting a B-chain
into 3 equal parts, a procedure which is attempted with probability
$n_{B'}/(n_A+n_B)$. One end of a B chain is chosen with probability
$1/2n_B$, and the cutting procedure starts from this end, leading to a
proposition probability: $P^{prop}_{B'\to kA} = 1/2(n_A+n_B)$.  For
the acceptance, one determines the possible number of bonding partners
$\nu_1$ and $\nu_2$ for the corresponding inverse move and accepts the
proposed move with probability: \[P^{acc}_{B' \to kA} =
\frac{\min(1,\exp(-\beta\Delta E(B'\to kA)+N_B\Delta\mu)}{\nu_1
\nu_2}\] Thus the choice of the acceptance probabilities fullfills the
detailed balance condition: \[P_{eq}(A) P^{prop}_{kA\to B'}
P^{acc}_{kA \to B'} = P_{eq}(B') P^{prop}_{B'\to kA} P^{acc}_{B' \to
kA}\]

\newpage

\begin{figure}[h]

\caption{The measured line of pseudo phase coexistence separating the
A-rich and A-poor phases, for which the concentration distribution has
two peaks of equal weight. The position of the critical point as
determined using the cumulant intersection method (see also
figure~\protect\ref{fig:cumulant}) is indicated, as are the measured
directions of the two relevant scaling fields \gM\ and \gE .}

\label{fig:cxcurve}
\end{figure}

\begin{figure}[h]

\caption{The value of the fourth order cumulant ratio $G_L=1-<m^4>/
3<m^2>^2$ with $m=\phi-\phi_{cx}$, expressed as a function of system
size $L$ and well depth $\epsilon$, along the line of pseudo phase
coexistence. An intersection occurs for a value of $G_L=0.47$ at
$\epsilon_c=0.02756(15),\Delta\mu_c=0.003603(15)$ }

\label{fig:cumulant}
\end{figure}

\begin{figure}[h]

\caption{The normalised concentration distribution for the $L=40$ and
$L=64$ system sizes obtained using the equal weight criterion with
$\phi^*=<\phi>$, at the assigned value of the critical well depth
$\epsilon=0.02756(15)$. The data are expressed in terms of the scaling
variable $x=a_{\cal M}^{-1} L^{\beta/\nu}(\phi-\phi_c)$, where the
non-universal scale factor $a_{\cal M}^{-1}$ has been chosen so that
distributions have unit variance. Also shown (solid) curve is the
fixed point function \ptMstar\ appropriate to the 3D Ising
universality class. Statistical errors do not exceed the symbol
sizes.}

\label{fig:ewr}
\end{figure}

\begin{figure}[h]

\caption{Extrapolation of $\phi_{cx}(L)$, defined in the text, against
$L^{(1-\alpha)/\nu}$. The least squares fit yields an infinite volume
estimate $\phi_c=0.03823(19)$. }

\label{fig:phic}
\end{figure}

\begin{figure}[h]

\caption{The measured antisymmetric field mixing corrections $\delta
p_L(x)$ of the $L=40$ and $L=64$ critical concentration distributions,
expressed in terms of the scaling variable $x=a_{\cal M}^{-1}
L^{\beta/\nu}(\phi-\phi_c)$ and shown as the data points. The data
has itself been corrected for a small off-coexistence correction as
described in the text. Also shown (full curve) is the universal
prediction following from equation~\protect\ref{eq:prediction},
utilising predetermined Ising forms \protect\cite{HILFER}}

\label{fig:asym}
\end{figure}

\begin{figure}[h]

\caption{The normalised distributions of the critical ordering
operator \pLM\ for the $L=40$ and $L=64$ system sizes, expressed as a
function of the scaling variable $y=a_{\cal M}^{-1}L^{\beta/\nu}({\cal
M}-{\cal M}_c)$. The full curve is the fixed point function \ptMstar\
appropriate to the Ising universality class \protect\cite{HILFER}. The
non-universal scale factor $a_{\cal M}^{-1}$ has been chosen so that
the distributions have unit variance. The data collapse corresponds to
a choice of the field mixing parameter $s=0.06(1)$. Statistical errors
do not exceed the symbol sizes.}

\label{fig:op}
\end{figure}

\begin{figure}[h]

\caption{The normalised distributions of the critical energy
density $p_L(u)$ for the $L=40$ and $L=64$ system sizes.}

\label{fig:en}
\end{figure}

\begin{figure}[h]


\caption{The normalised distribution of the energy-like operator \pLE\
(cf. equation~\protect\ref{eq:opscale4}) expressed as a function of
the scaling variable $z=a_{\cal E}^{-1}L^{(1-\alpha)/\nu}({\cal
E}-{\cal E}_c)$ for the $L=40$ and $L=64$ system sizes. The full curve
is the fixed point function \ptEstar\ appropriate to the Ising
universality class \protect\cite{HILFER}. The non-universal scale
factor $a_{\cal E}^{-1}$, has been chosen so that the distributions
have unit variance. The data collapse shown corresponds to a choice of
the field mixing parameter $r=-1.00(3)$. Statistical errors do not
exceed the symbol sizes.}

\label{fig:enop}
\end{figure}

\begin{figure}[h]

\caption{The finite size scaling behavior of the variance of the critical
energy, energy operator and ordering operator distributions,
c.f. figures~\protect\ref{fig:en} and~\protect\ref{fig:enop}. The
straight lines superimposed on the data points have the forms
$0.0076L^{\gamma/\nu}$ (solid line) and $2.7L^{\alpha/\nu}$ (broken
line), where $\gamma/\nu=1.970, \alpha/\nu=0.211$.}

\label{fig:cv}
\end{figure}

\begin{figure}[h]
\caption{Illustration of the semi-grandcanonical Monte-Carlo moves for $k=3$
and $N_B=3$.
	 Arrows indicate the bonds that are removed when one creates $k=3$ A-chains or
	 which have to be appropriately added in the inverse case. The monomer
positions
	 are left unaltered.}

\label{fig:algo}
\end{figure}

\end{document}